\documentclass[3p,times,procedia]{elsarticle}
\usepackage{nupha_ecrc}

\volume{00}

\firstpage{1}

\journalname{Nuclear Physics A}

\runauth{}

\jid{nupha}

\jnltitlelogo{Nuclear Physics A}

\usepackage{amssymb}

\usepackage[figuresright]{rotating}
\usepackage{subcaption}
\usepackage{lineno}

   \newcommand{\Dzero}{D$^{0}$}
   \newcommand{\Dplus}{D$^{+}$}
   \newcommand{\Dstar}{D$^{*+}$}
   \newcommand{\Dsubs}{D$_{\mathrm{s}}^{+}$}
   \newcommand{\pbpb}{Pb--Pb}
   \newcommand{\sNN}{$\sqrt{s_{\rm {NN}}}$}

\begin{document}

\begin{frontmatter}




\title{D-meson production in Pb--Pb collisions with ALICE at the LHC}


\author{Stefano Trogolo \\ \normalsize{on behalf of the ALICE Collaboration}}

\address{Universit\`a di Padova, D.F.A. and INFN, Sez. Padova - Via Marzolo 8, 35131 Padova - Italy}

\begin{abstract}
Heavy quarks, like charm and beauty, are sensitive probes to investigate the colour-deconfined medium created in high-energy heavy-ion collisions, the Quark--Gluon Plasma (QGP). The ALICE Collaboration measured the production of \Dzero, \Dplus, \Dstar and \Dsubs\ in \pbpb\ collisions at \sNN\ = 5.02 TeV. The measurement of the nuclear modification factor ($R_{\rm {AA}}$) provides information on the in-medium parton energy loss. In addition, the comparison between \Dsubs\ and the non-strange D-meson $R_{\rm {AA}}$ allows to investigate possible modifications of the charm-quark hadronisation mechanism in the QGP.
The most recent results for these observables, which were obtained by analysing the latest 2018 data sample of \pbpb\ collisions as well as the comparison with theoretical model calculations, are presented. 
\end{abstract}


\end{frontmatter}



\section{Introduction}
Heavy quarks (i.e. c and b quarks) are excellent probes for the characterisation of the deconfined medium created in ultra-relativistic heavy-ion collisions, the Quark--Gluon Plasma (QGP). Because of their large mass, heavy quarks are mainly produced in the early times of the collisions, before the formation of the QGP, via hard-scattering processes \cite{Prino:2016cni}. 
The nuclear modification factor ($R_{\rm {AA}}$) of hadrons containing heavy quarks, which is defined as the ratio between the transverse momentum ($p_{\rm {T}}$) spectrum in nucleus--nucleus collisions (d$N_{\rm {AA}}$/d$p_{\rm {T}}$) and the $p_{\rm {T}}$-differential cross section measured in pp collisions (d$\sigma_{\rm {pp}}$/d$p_{\rm {T}}$) scaled by the average nuclear overlap function $\langle T_{\rm {AA}}\rangle$, is used to inquire the properties of the in-medium parton energy loss. The comparison between light-flavour and heavy-flavour hadrons gives insight into the dependence of the energy loss on the colour charge and quark mass \cite{Prino:2016cni,Andronic:2015wma,Djordjevic}.
Furthemore, it is predicted that heavy quarks could hadronise via coalescence in the medium and, therefore, the comparison of the heavy-flavour hadrons with and without strangeness can provide information on the possible modifications of the hadronisation mechanism \cite{Rapp:2018qla}.

Charmed mesons are reconstructed in ALICE \cite{jinst:s08002, ijmo:s08002} at mid rapidity ($|\textit{y}|<$ 0.5) via the following decay channels: \Dzero\ $\rightarrow$ $\rm {K^-\pi^+}$ (BR = 3.95$\%$), \Dplus\ $\rightarrow \rm {K^-\pi^+\pi^+}$ (BR = 9.46$\%$), \Dstar\ $\rightarrow$ \Dzero\ $\rm {\pi^+} \rightarrow \rm {K^-\pi^+\pi^+}$ (BR = 2.63$\%$) and \Dsubs\ $\rightarrow\rm {\phi\pi^+} \rightarrow \rm {K^-K^+\pi^+}$ (BR = 2.27$\%$) and their charge conjugates \cite{PhysRevD.98.030001}. Kaons and pions are identified with the Time Projection Chamber \cite{Alme:2010ke} via their specific energy loss and with the Time-Of-Flight detector \cite{Akindinov:2013ke}. Particle identification (PID) of the decay products and geometrical selections on the displaced decay-vertex topology are applied to reduce the combinatorial background. The D-meson raw yields are extracted with an invariant-mass analysis as described in \cite{Acharya2018}. The efficiency and acceptance corrections are obtained from MC simulations based on HIJING \cite{hijing} event generator and enriched with PYTHIA \cite{pythia6} events with c$\bar{\rm {c}}$ and b$\bar{\rm {b}}$ pairs. The fraction of prompt D mesons is estimated with a FONLL-based approach \cite{fonll, Acharya2018}.

This contribution presents the most recent results on the D-meson production measured by analysing the 2018 data sample of \pbpb\ collisions at a centre-of-mass energy per nucleon pair of \sNN\ = 5.02 TeV, which corresponds to an integrated luminosity of about 114 $\mu$b$^{-1}$ (49 $\mu$b$^{-1}$) for the 0--10$\%$ (30--50$\%$) centrality class. The hadronic cross section used for the luminosity calculation is taken from \cite{PhysRevC.97.054910}. 
Here, the cross section of D-meson production in pp collisions, collected during the 2017 pp run at $\sqrt{s}$ = 5.02 TeV, is taken from the published results \cite{Acharya2019}.

\section{Non-strange D-meson nuclear modification factor}
\begin{figure*}[t!]
	\centering
	\begin{subfigure}[t]{0.49\textwidth}
		\centering
	    \includegraphics[scale=0.1]{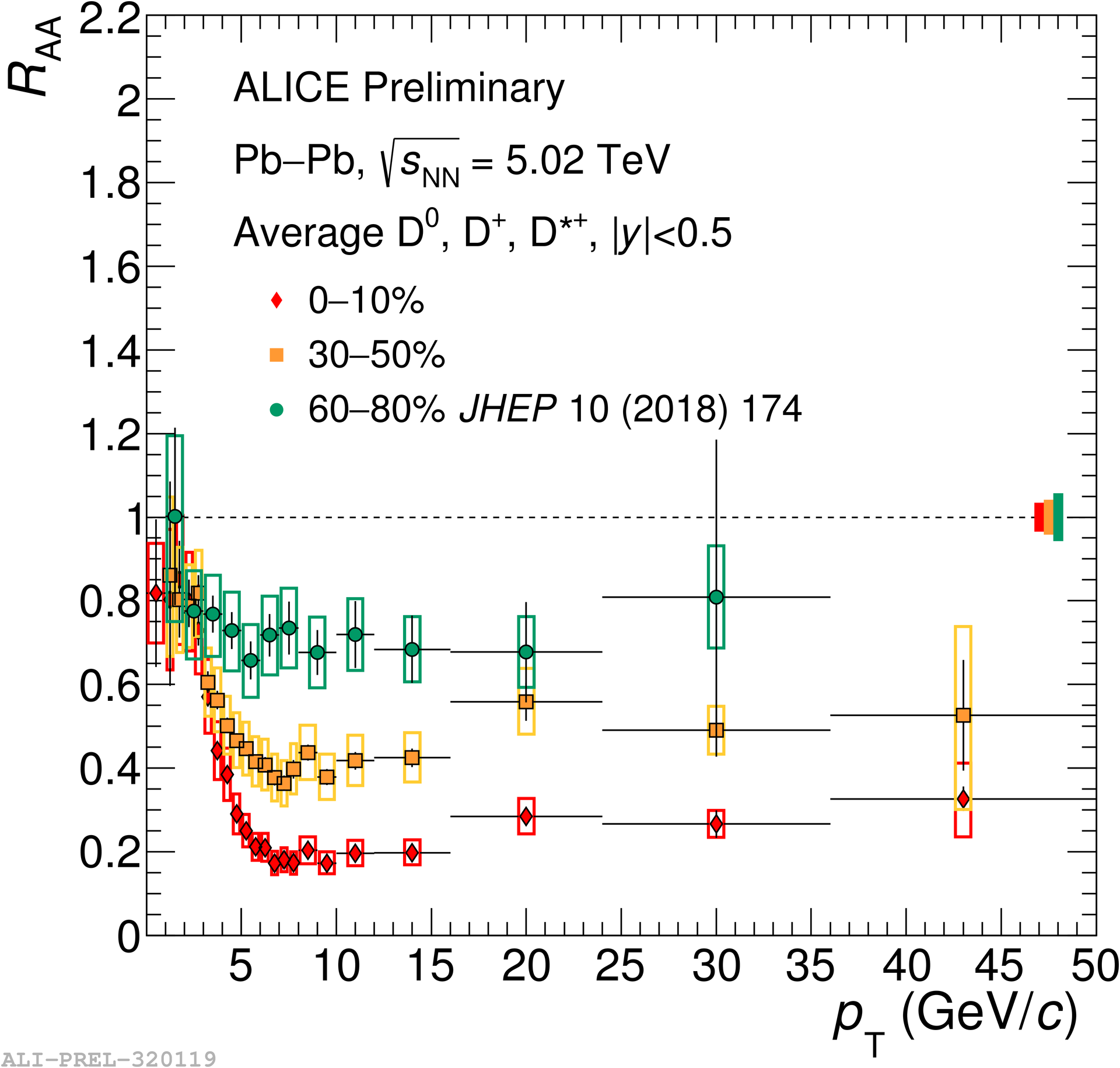}
    \end{subfigure}
    \begin{subfigure}[t]{0.49\textwidth}
    	\centering
	    \includegraphics[scale=0.125]{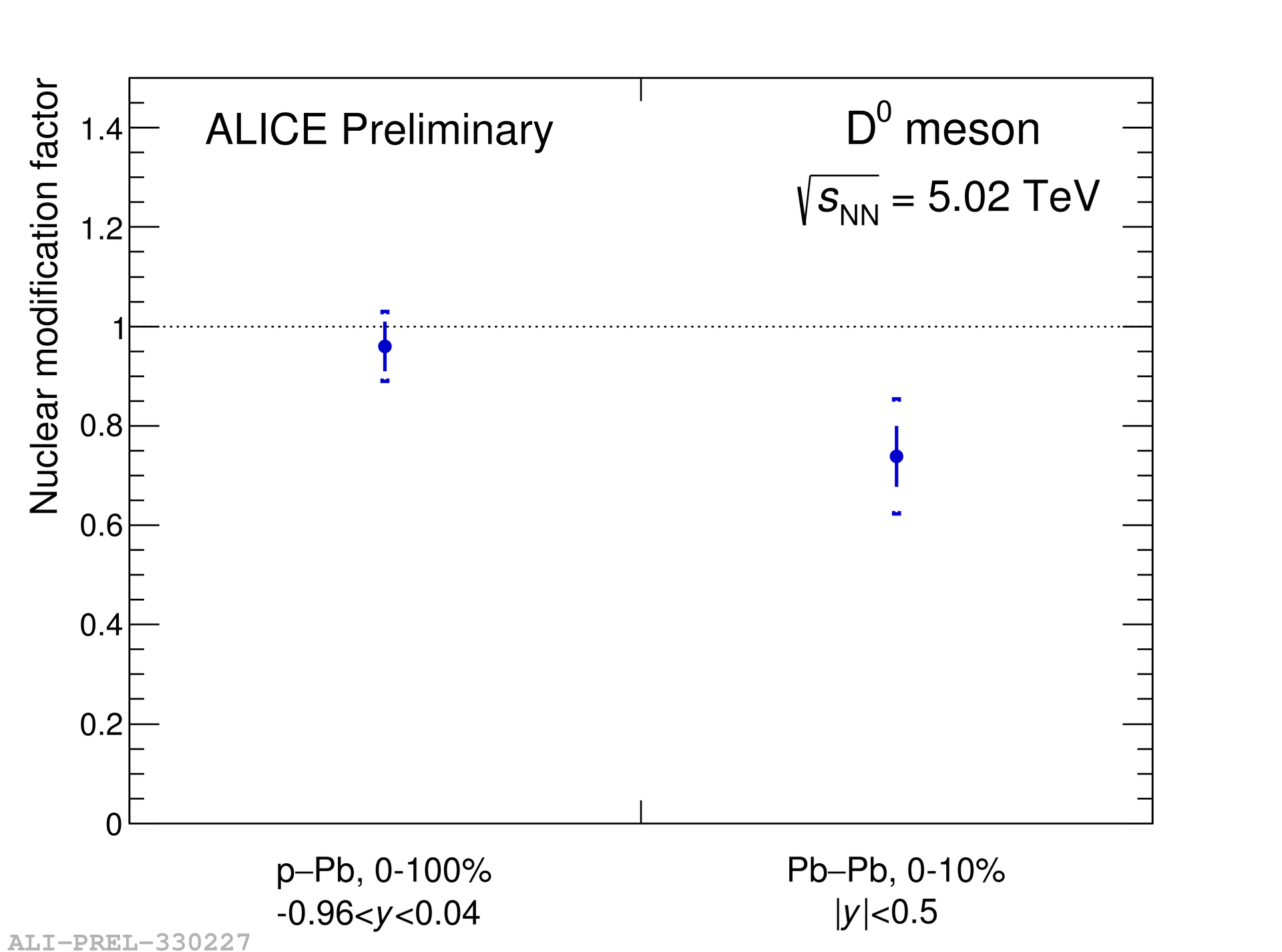}
    \end{subfigure}
    \caption{(left panel) Average $R_{\rm {AA}}$ of prompt \Dzero, \Dplus\ and \Dstar\ in the 0--10$\%$, 30--50$\%$ and 60--80$\%$ \cite{Acharya2018} centrality class of \pbpb\ collisions at \sNN\ = 5.02 TeV; (right panel) $p_{\rm {T}}$-integrated prompt \Dzero\ $R_{\rm {AA}}$ in \pbpb\ collisions at \sNN\ = 5.02 TeV compared to the results measured in p--Pb collisions at the same centre-of-mass energy \cite{PhysRevC.94.054908}.}
    \label{nonstrange_raa_centrality}
\end{figure*}

The $R_{\rm {AA}}$ of prompt \Dzero, \Dplus, and \Dstar\ is measured in central (0--10$\%$) and semi-central (30--50$\%$) and compared with the result in peripheral (60--80$\%$) collisions previously published (2015 data sample) \cite{Acharya2018}. Figure \ref{nonstrange_raa_centrality} (left panel) shows an increasing suppression from peripheral to central collisions compared to pp collisions \cite{Acharya2019}.

Furthermore, the first measurement of the \Dzero\ down to $p_{\rm {T}}$ = 0 in \pbpb\ collisions is performed by exploiting only the PID capabilities of the ALICE detectors.
Thus, the $p_{\rm {T}}$-integrated $R_{\rm {AA}}$ of the \Dzero\ is computed using the measured $R_{\rm {AA}}$ and
an FONLL-based calculation for the extrapolation in the unmeasured $p_{\rm {T}}$ regions, that are below 1 GeV/$c$ in p--Pb and above 36 GeV/$c$ (50 GeV/$c$) in p--Pb (Pb--Pb) collisions.
Figure \ref{nonstrange_raa_centrality} (right panel) shows the result obtained in \pbpb\ collsions, which is significantly below unity, compared with the one measured in p--Pb collisions at the same  centre-of-mass energy \cite{PhysRevC.94.054908}.

\begin{figure*}[t!]
    \centering
    \begin{subfigure}[t]{0.49\textwidth}
        \centering
        \includegraphics[scale=0.10]{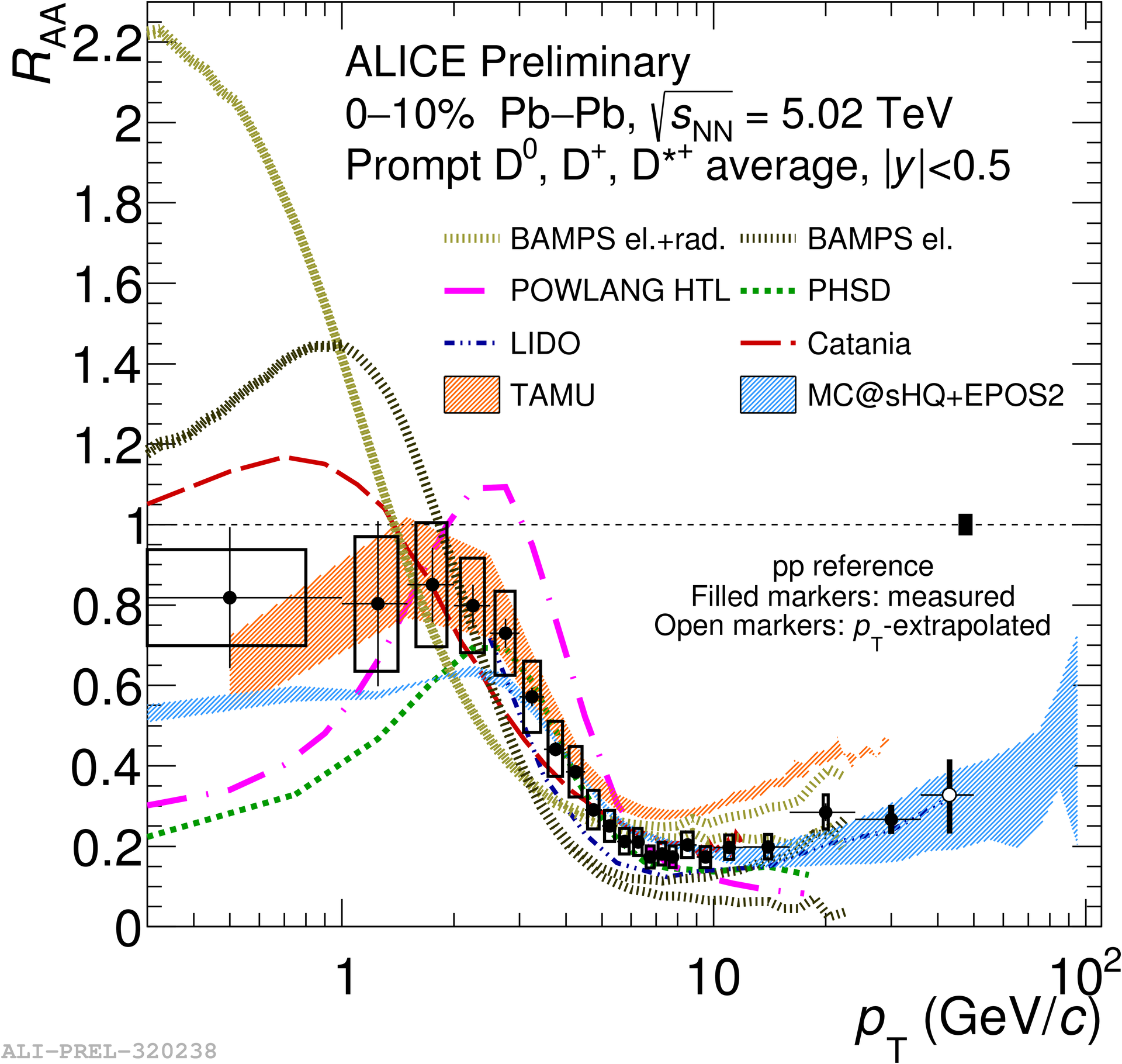}
    \end{subfigure}
    \begin{subfigure}[t]{0.49\textwidth}
        \centering
        \includegraphics[scale=0.10]{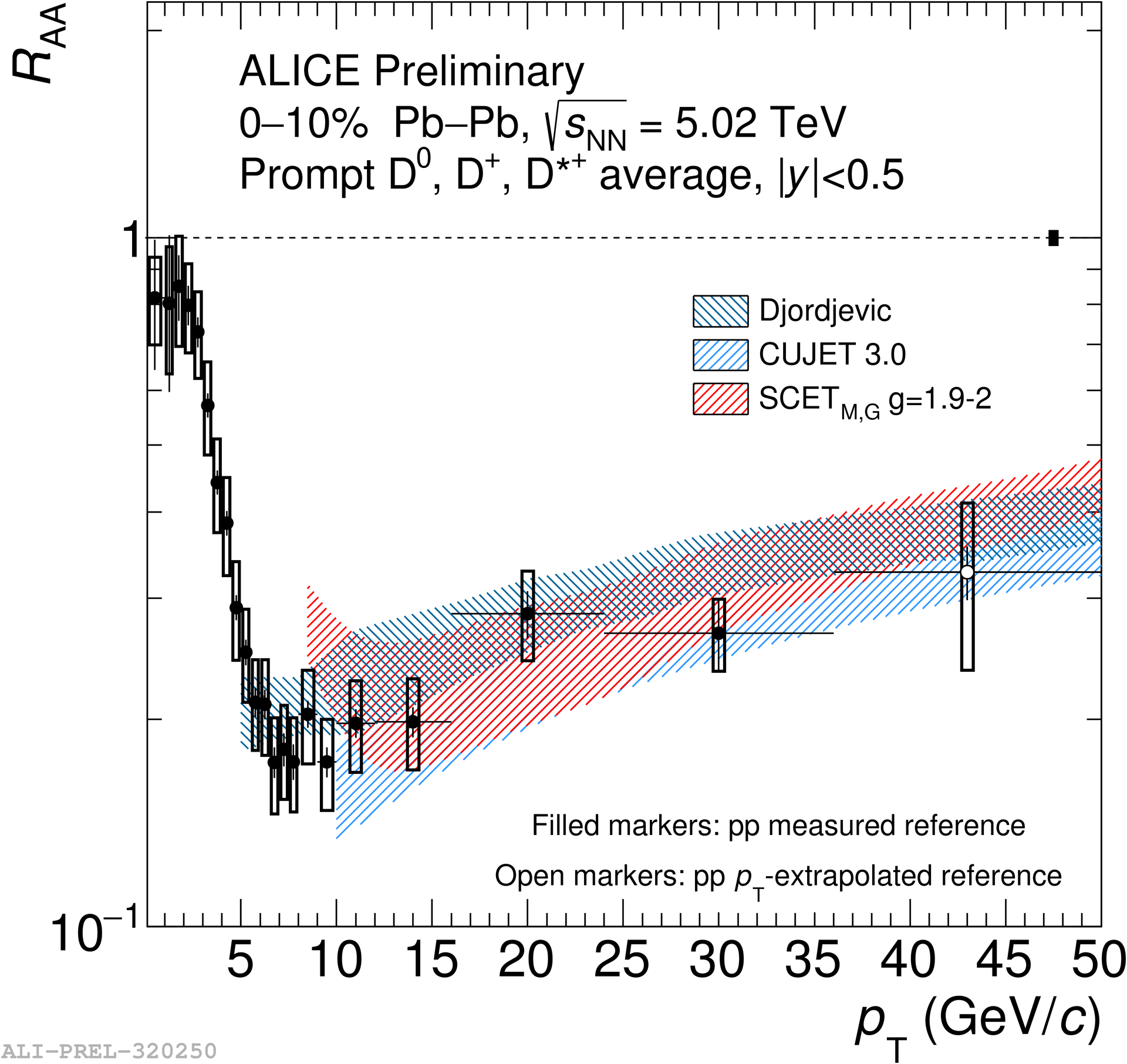}
    \end{subfigure}
    \caption{Average prompt \Dzero, \Dplus, and \Dstar\ $R_{\rm {AA}}$ in \pbpb\ collisions at \sNN\ = 5.02 TeV in the 0--10$\%$ centrality class compared to model predictions based on the charm-quark transport in an expanding medium (left panel) and on pQCD calculations of parton energy loss (right panel).}
    \label{nonstrange_raa_models}
\end{figure*}

The left panel of Fig. \ref{nonstrange_raa_models} shows the average prompt non-strange D-meson $R_{\rm {AA}}$ compared to the predictions by theoretical models focused on the collisional energy loss in an hydrodynamically expanding medium and recombination effect on hadronization \cite{BAMPS, POWLANG, LIDO, TAMU, PHSD, Catania, Gossiaux2014}. On the other hand, the comparison to the predictions, obtained via perturbative QCD (pQCD) calculations, by models that describes the radiative energy loss and the hadronization via fragmentation of high $p_{\rm {T}}$ charm-quark \cite{Djordjevic, CUJET, SCET} is reported in the right panel.
The $R_{\rm {AA}}$ for $p_{\rm {T}}<$ 10 GeV/$c$ is fairly well reproduced by most of the models that implement heavy-quark transport in the medium with a realistic hydrodynamical evolution. The high-$p_{\rm {T}}$ region of the $R_{\rm {AA}}$ is well described by pQCD-based models, as well as by the MC@sHQ+EPOS2 and LIDO models.   

\begin{figure*}[b!]
	\centering
    \includegraphics[scale=0.10]{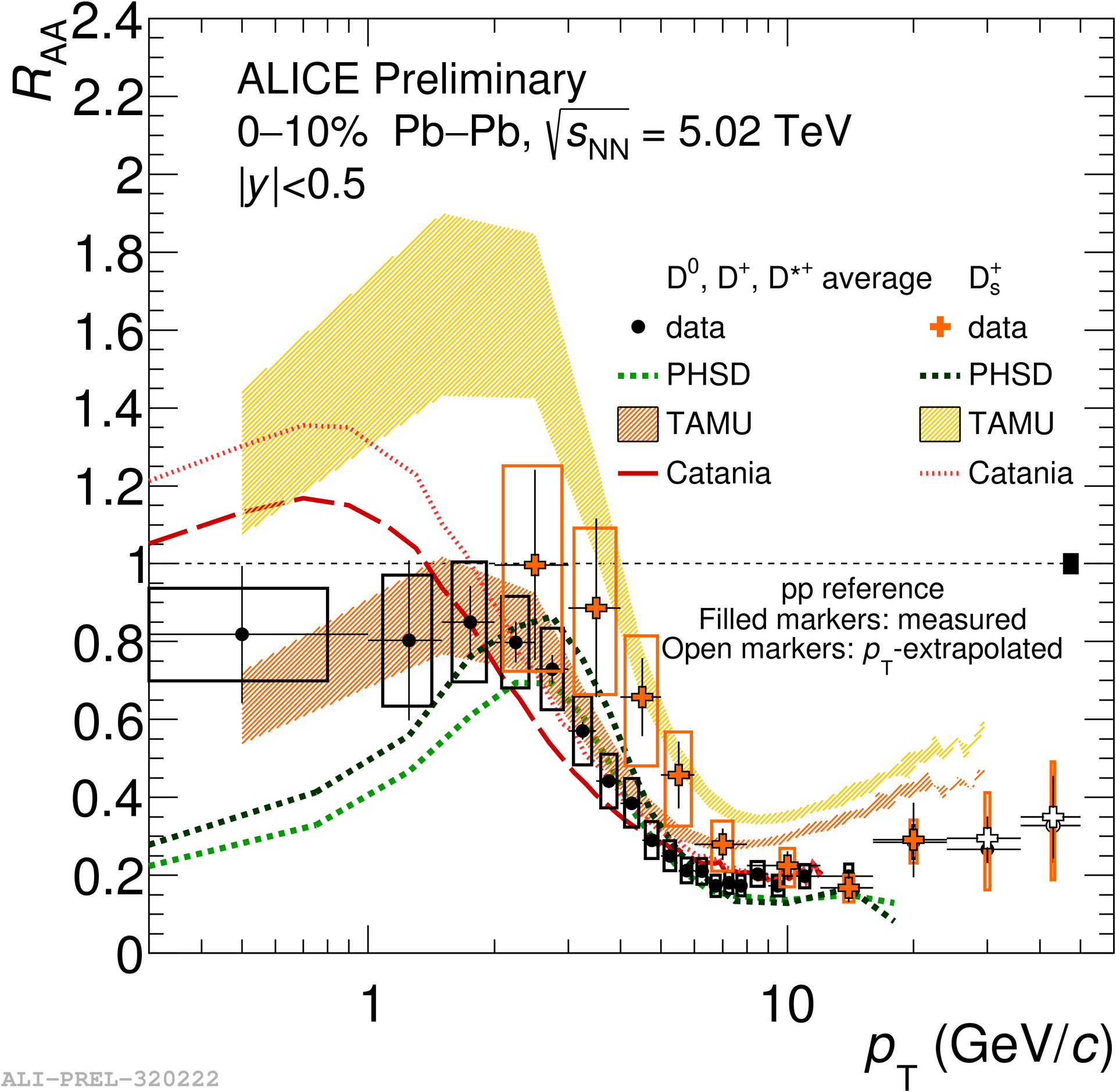}
    \caption{Average $R_{\rm {AA}}$ of non-strange D mesons and $R_{\rm {AA}}$ of \Dsubs\ meson in the 0--10$\%$ class compared with heavy-quark transport models \cite{TAMU, PHSD, Catania}.} 
    \label{strange_raa_models}
\end{figure*}

\section{\Dsubs\ nuclear modification factor}
\begin{figure*}[t!]
	\centering
    \includegraphics[scale=0.21]{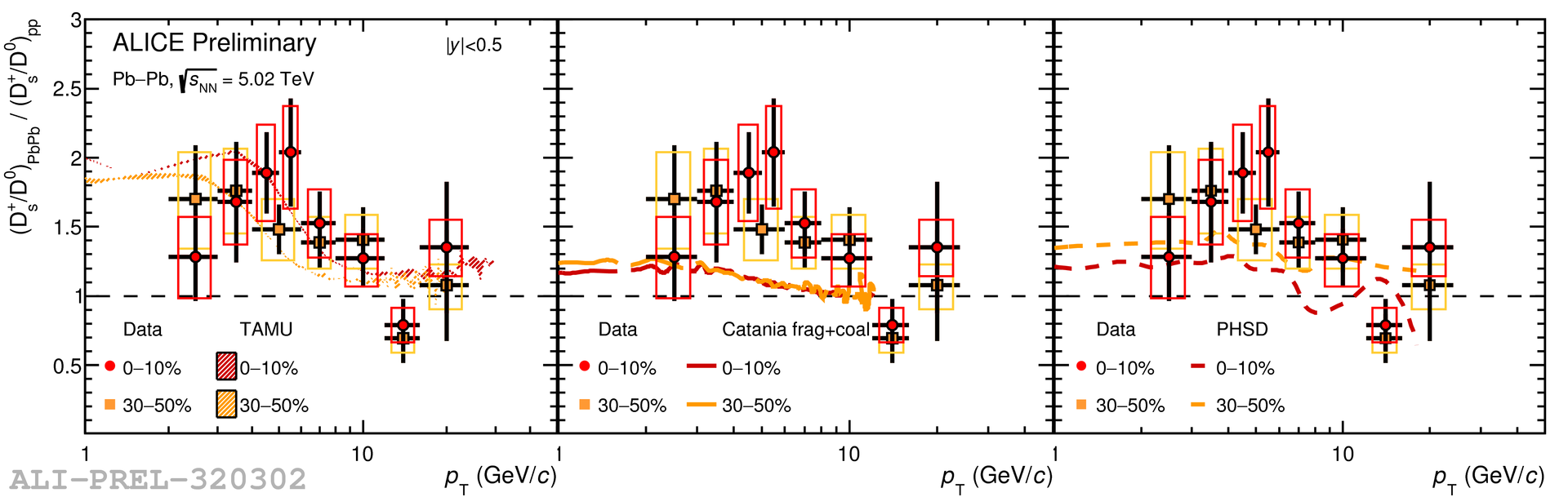}
    \caption{Ratio of \Dsubs/\Dzero\ measured in 0--10$\%$ and 30--50$\%$ central \pbpb\ collisions at \sNN\ = 5.02 TeV over the measurement done in pp collisions at $\sqrt{s}$ = 5.02 TeV \cite{Acharya2019} compared with transport models \cite{TAMU, PHSD, Catania}.}
    \label{double_DSoverDzero}
\end{figure*}

The \Dsubs\ $R_{\rm {AA}}$ is measured in the same centrality classes of non-strange D mesons, using the 2018 \pbpb\ dataset, to investigate possible differences in the charm-quark hadronisation mechanism. A supervised Machine Learning (ML) technique is adopted for the optimization of the signal extraction which allows the extension of the low and high $p_{\rm {T}}$ reach of the measurement. The algorithm used is a Boosted Decision Tree (BDT) and it is applied on variables related to the PID of the decay products and the decay topology. The training is done using signal and background taken from a MC simulation and a fraction of the real data, respectively. 
Figure \ref{strange_raa_models} shows the prompt  \Dsubs\ $R_{\rm {AA}}$ compared to the non-strange D-meson $R_{\rm {AA}}$ and to the theoretical predictions of models based on heavy-quark transport in the medium and coalescence. The comparison shows a hint of a \Dsubs\ $R_{\rm {AA}}$ larger than that of non-strange D mesons for $p_{\rm {T}}<$ 10 GeV/$c$, as expected in case of hadronisation via coalescence due to the enhanced production of strange quarks in the QGP. The enhancement of \Dsubs\ over non-strange D mesons is highlighted in Fig. \ref{double_DSoverDzero} where the ratio between \Dsubs/\Dzero\ measured in \pbpb\ collisions and the result obtained in pp collisions is shown and compared with the theoretical calculations based on transport models. The ratio is fairly described by models including interactions with only collisional processes \cite{TAMU}.

\section{Conclusions}
The measurements of \Dzero, \Dplus, \Dstar, and \Dsubs\ $R_{\rm {AA}}$, performed on the latest sample of \pbpb\ collisions at \sNN\ = 5.02 TeV collected in 2018, show an improved statistical precision by about a factor three (two) in the central (semi-central) collisions with respect to the previous results published by the ALICE Collaboration \cite{Acharya2018}.  
During the LHC Run 3, ALICE is expected to perform even more precise measurements thanks to the improved precision of the upgraded detectors and to the larger data sample that will be collected.





\bibliographystyle{elsarticle-num}
\bibliography{biblio.bib}







\end{document}